\documentclass[11pt,preprint]{aastex}

\begin{document}

\noindent
{\LARGE{\bf How to collect matches that will catch fire}}

\vskip 0.3in

\noindent
{\Large Abraham Loeb}

\vskip 0.25in

\noindent
{\Large{\bf Science can only blossom if young researchers are rewarded for
  acquired skills and growth rather than inherited academic ancestry.}}

\vskip 0.35in

\noindent
{\huge W}ill a match catch fire when it scratches against the rugged
matchbox wall?  Knowing the answer is of paramount importance if we
want to collect useful matches in our box. One way to find out is to
try them all. The only problem with this approach is that by the time
we will know the answer, the burnt matches will be of no value. The
challenge is how to select useful matches reliably in advance?
Putting this challenge into an academic context, {\it how can we
  select a cohort of promising scientists before they have made their
  discoveries?}  This is the fundamental challenge of academic
planning.  Prestigious universities are plagued by past hirings which
led to `duds' or `dead wood', namely faculty who when hired were
labeled as geniuses with great promise but in retrospect, decades
later, had little impact on the progress of science. At the same time,
some of their contemporaries who were not endorsed by prominent
scientists and hence moved to faculty positions at lesser schools,
carried the day. Without mentioning names, suffice it to say that this
is a familiar occurrence. Why is this phenomenon so prevalent?

Senior scientists who serve on promotion, prize, or search committees
are often asked to evaluate the promise of their younger
colleagues. One would naively expect them to approach this challenge
in the same way that they address a scientific problem, namely study
all available data and construct a model that extrapolates into the
future. In order to avoid biases, it would appear natural to adopt a
dynamical model which takes into consideration the special
initial conditions of an individual and allows for growth in
forecasting the individual's future. For example, a young researcher
who did not benefit from the privilege of being nurtured by top
quality mentors or had to transition from a different culture or an
inferior socio-economic status, should be given more slack. This is
common sense. But is it common practice?

My experience over the past three decades suggests otherwise. Young
scientists are commonly assigned static labels without proper
attention to their starting point or the time derivative of their
career trajectory.  Early-career evaluations reflect a frozen snapshot
of achievements. It is common occurrence for a science department to
under-appreciate a faculty position applicant who graduated many years
ago from the same department, due to a frozen image of the
qualifications of the applicant. These mistakes have serious
consequences, as poor recruitments lead to drifts in the prestige of
academic institutions.  To make things worse, evaluators often resist
updating their image of an individual later on, out of fear that
admitting the need for this update would reflect badly as lack of
foresight originally.

Insisting on a static image that is out of synch with the growth of a
successful researcher often leads to persistent attempts to shape
reality in such a way that it will justify the pre-conception. The
inconvenient truth is that evaluators with pre-conceptions have the
power to allocate resources so as to justify their original static
images.  On the positive side, when serving on prize committees they
can favor those whom they originally supported. On the negative, when
serving on a grant allocation committee, they could block support for
others even in the face of evidence that contradicts their early
impressions. Such actions lead to self-confirming prophecies, and can
occasionally crash the rising career of brilliant individuals who were
not recognized as such at an earlier stage of their career.

The above faults are sometimes driven by the misconception that
scientific success is enabled primarily by raw talent that should be
evident in any snapshot of an individual. After all, Albert Einstein
showed brilliance at a very young age. But this presumes a static view
of science itself, while in reality the landscape of science has
evolved dramatically over the past century since the days of
Einstein. Today much of the relevant scientific information is
updating rapidly and there are many more scientists around. In this
climate, success is often linked to acquired skills, such as the
abilities to adjust to rapidly changing intellectual landscapes (e.g.,
Big Data) and to identify the right problem to work on while others
are searching in the dark. Today's science requires social skills,
namely the ability to lead other scientists and to communicate results
so that they promote progress. These acquired skills take time to
develop and require any model that attempts to forecast success
reliably to include evolution and refrain from static images.

Yet, it often seems that the guiding principles are completely off
target.  A traditional obstacle to an honest evaluation process is
that prominent scientists wish to promote their own research program
in an effort to link it permanently to the mainstream.  Frequently,
this tendency takes the form of senior scientists promoting their own
students or group members well beyond what may count as fair
play. This tends to suppress independent thinking outside the
boundaries of widely-held paradigms.  Put simply, senior scientists
tend to measure success by how much a younger colleague replicates
their own research agenda or set of skills. For example, if they are
fluent with mathematical subtleties, they will identify success with
mathematical skills. In faculty recruitments, this tendency for
self-replication is dangerous because it might not stop at academic
qualifications, but could easily spill over to an unconscious bias
based on the replication of one's own gender, race or ethnicity.

There are multiple paths to success in today's science. Some paths are
mathematical and quantitative while others are qualitative and require
conceptual vision. Rather than replicating ourselves and preserving a
static past, we would do better to aim at diversity and promote
scientists of all varieties to secure a vibrant future. When serving
on committees, we should resist static images of our younger
colleagues and replace them with dynamical models by paying special
attention to initial conditions and embracing evolution in our
assessments. To cultivate innovation, we should always encourage
creativity beyond the comfort limits that we establish for
ourselves. Keeping a wide variety of matches in our matchbox will
guarantee that not all of them will be duds. Hopefully, a few will
light up in the dark to guide us how to move forward.

\bigskip
\bigskip
\bigskip

\noindent{\it Abraham Loeb is the Frank B. Baird Jr. Professor of Science
  at Harvard University. He serves as chair of the Harvard Astronomy
  department and director of the Institute for Theory \& Computation
  at the Harvard-Smithsonian Center for Astrophysics, 60 Garden
  Street, Cambridge, Massachusetts 02138, USA. \\e-mail:
  aloeb@cfa.harvard.edu}

\acknowledgements 

\end{document}